\documentclass[]{aa}  
\usepackage{natbib}
\bibpunct{(}{)}{;}{a}{}{,} 
\usepackage{hyperref} 
\usepackage{graphicx}
\usepackage{txfonts}
\begin{document} 
\title{
The effects of a revised $^7$Be e$^-$-capture rate on solar neutrino fluxes}
%
%
\author{D. Vescovi \inst{1,2}
\and L. Piersanti \inst{3,2}
\and S. Cristallo \inst{3,2}
\and M. Busso \inst{4,2}
\and F. Vissani \inst{5}
\and S. Palmerini \inst{4,2}
\and S. Simonucci \inst{6,2}
\and S. Taioli \inst{7,8}
}

\institute{Gran Sasso Science Institute, Viale Francesco Crispi, 7, 67100 L'Aquila, Italy\\
\email{diego.vescovi@gssi.it}
\and INFN, Section of Perugia, Via A. Pascoli snc, 06123 Perugia, Italy
\and INAF, Observatory of Abruzzo, Via Mentore Maggini snc, 64100 Teramo, Italy
\and University of Perugia, Department of Physics and Geology, Via A. Pascoli snc, 06123 perugia, Italy
\and INFN, Laboratori Nazionali del Gran Sasso, Via G. Acitelli, 22, Assergi, L’Aquila, Italy
\and Division of physics School of Science and Technology Università di Camerino, Italy
\and Faculty of Mathematics and Physics, Charles University, Prague, Czech Republic
\and European Centre for Theoretical Studies in Nuclear Physics and Related Areas (ECT*-FBK) and Trento Institute for Fundamental Physics and Applications (TIFPA-INFN), Trento, Italy
}

\date{Received ; accepted }

 
  \abstract
   {The electron-capture rate on $^7$Be is the main production channel for $^7$Li in several astrophysical environments. Theoretical evaluations have to account for not only the nuclear interaction, but also the processes in the plasma where $^7$Be ions and electrons interact. In the past decades several estimates were presented, pointing out that the theoretical uncertainty in the rate is in general of few percents.}
   {In the framework of fundamental solar physics, we consider here a recent evaluation for the $^7$Be+e$^-$ rate, not used up to now in the estimate of neutrino fluxes.}
   {We analysed the effects of the new assumptions on Standard Solar Models (SSMs) and compared the results obtained by adopting the revised $^7$Be+e$^-$ rate to those obtained by the one reported in a widely used compilation of reaction rates (ADE11).}
   {We found that new SSMs yield a maximum difference in the efficiency of the $^7$Be channel of about -4\% with respect to what is obtained with the previously adopted rate. This fact affects the production of neutrinos from $^8$B, increasing the relative flux up to a maximum of 2.7\%. Negligible variations are found for the physical and chemical properties of the computed solar models.}
   {The agreement with the SNO measurements of the neutral current component of the $^8$B neutrino flux is improved. }


\maketitle
%
%
\section{Introduction}
\label{sec:intro}
Solar models and their comparisons with observations are a powerful tool for probing the solar interiors with high accuracy, describing the trend of the sound speed and predicting how neutrinos are distributed among the various channels \citep[see e.g.][ for a review]{bahc01}.

Solar neutrino measurements, in particular those from the $^8$B channel \citep{ahar13,abe16} yielded information on fundamental neutrino properties;  nowadays these properties are measured with an increasing accuracy  and detailed knowledge of neutrino fluxes maintains its importance also for this aim.

Very recently the Borexino collaboration presented the first global analysis of three individual neutrino components of the proton-proton (pp) chain, namely pp, $^7$Be and pep neutrinos, putting also an upper limit to those from CNO, over an energy range from 0.19 MeV to 2.93 MeV \citep{agos18}. 

These new data on neutrino fluxes can be used to improve our knowledge of the solar interiors \citep{viny17}, which is still beset with problems; among them, of special relevance are those raised by the compilations of solar abundances based on 3D atmospheric models \citep{asplund05}, which lead to disagreements with the measured sound speed \citep{bahc05b}.

Standard solar model predictions for neutrino fluxes are then very sensitive to the reaction rates adopted, obviously including electron-captures in the plasma (which are also of great importance for several other astrophysical problems).  The electron-capture rate on $^7$Be itself is strongly dependent on the density and temperature distribution in the stellar structure \citep{simo13}; in solar conditions, in particular, this destruction channel of $^7$Be dominates over proton captures \citep{adel98}.  From this latter branching, through $^8$B-decays, further neutrinos are emitted and can be detected by experiments like Super-Kamiokande, SNO and KamLand. The observed flux of $^8$B neutrinos is expected to be inversely proportional to the electron-capture rate on $^7$Be, being the counting rate in experiments determined by the number of proton-capture reactions occurring per
unit of time \citep{bahc69}.
Despite many different estimates presented \citep{bahc62,bahc69,john92,gruz97}, the accuracy in our knowledge of the relative importance of these two channels in not yet satisfactory and improvements have been limited over the years. 

In this work we make a step forward by using a new estimate of the electron-capture rate on $^7$Be \citep[][ hereafter STPB13]{simo13} to compute SSMs.
The results are then compared  with those obtained by the widely used rate by \citet{adel11} (hereafter ADE11), focusing our attention on the solar neutrino fluxes.
We make use of a tabulated version of the  decay rate by STPB13. The aforementioned table, available at the CDS, contains the following information. Column 1 lists the density over the mean molecular weight for electrons in units of ${\rm g \, cm^{-3}}$, Column 2 gives the temperature in units of ${\rm K}$ and Column 3 provides the value of the electron-capture rate in units of ${\rm s^{-1}}$. All the quantities are expressed in logarithmic scale. 
We also present an analytical approximation to it (see section \ref{sec:capture}).
Our work is organized as follows. In Section \ref{sec:ssm} the main features of the adopted stellar evolutionary code and of SSMs are described. 
Section \ref{sec:capture} illustrates the calculation of the electron-capture rate on $^7$Be and presents a comparison with the previous estimate. 
In Section \ref{sec:fluxes} we analyze the main characteristics of the ensuing SSM, while in Section \ref{sec:impact} the impact of the adopted rate on neutrinos from the $^8$B channel is discussed.
We summarize our results in Section \ref{sec:concl}.
\section{The Standard Solar Model}
\label{sec:ssm}
A SSM represents the mathematical way of fitting the present-day Sun status, provided some boundary conditions as luminosity, radius, mass and composition are available. Other important features such as temperature, pressure, sound-speed profiles, solar photospheric abundances and neutrino fluxes can then be predicted. Each of these quantities strictly depends on the nuclear reactions at work in the Sun's interiors, whose main outcome is helium production by hydrogen burning. This occurs through the pp-chain ($\sim$99$\%$) and, to a much lesser extent, through the CN-cycle ($\sim$1$\%$). Although the latter is not very important for the energy production in our Sun, it is relevant for the details of the neutrino production and as a test of the correctness of the predictions. 
Other ingredients of the input physics, such as equation of state (EoS), opacity, chemical composition, etc. are also crucial to predict the solar quantities mentioned above.

The essentials of a SSM include the full evolution of a 1 $M_{\sun}$ star from the pre-main sequence to the present solar age $t_\sun$ = 4.566 Gyr, usually by considering that mass-loss is negligible. In addition, a SSM is required to reproduce, once  the presolar composition is fixed, the present-day solar mass $M_{\sun}$, age, radius $R_{\sun}$, and luminosity $L_{\sun}$ as well as the observed metal-to-hydrogen ratio $(Z/X)_\sun$ at the surface of the Sun.
In order to do this, in our models we calibrated accordingly, with an iterative procedure, the initial helium and the metal mass fractions $Y_{\rm ini}$ and $Z_{\rm ini}$, respectively) as well as the mixing-length parameter ($\alpha_{\rm MLT}$).
Our solar models have been calculated with the FUNS stellar evolutionary code \citep{stra06,pier07,cris11}.
All the models assume a present solar luminosity of $L_{\sun} = 3.8418 \times 10^{33}$ erg s$^{-1}$, a present solar radius $R_{\sun} = 6.9598 \times 10^{10}$ cm and a solar mass $M_{\sun} = 1.989 \times 10^{33}$ g \citep{alle63,bahc05a}.

The input physics is basically the same adopted by \citet{pier07}, but includes a few recent updates as listed below.
We adopted the nuclear reaction rates presented in Table \ref{tab:rates}, except for the case of the $^7$Be electron-captures, for which we used either the rate suggested by \citet{adel11} or the one computed by \citet{simo13}. 
Concerning the mean energy loss in the individual branches of neutrino production, we used the experimental values 
suggested by \citet{viss18} (see their Table 2).
For electron screening effects in the solar plasma we adopted the Salpeter formula for the weak-screening, as recommended by \citet{gruz98} and \citet{bahc02}.
The EoS is the same as the one described by \citet{stra88} for fully ionized matter, in the form updated by \citet{prad02} for log\textit{T} [K] $\geq$ 6.0 and a Saha equation for log\textit{T} [K] $<$ 6.0.
Atomic diffusion has been included, taking into account the effects of gravitational settling and thermal diffusion, by inverting the coupled set of Burgers equations \citep{thou94,pier07}.
For radiative opacities, we used the OPAL tables \citep{igle96} for high temperatures (log\textit{T} [K] $\geq$ 4.0) and the \citet{ferg05} molecular opacities for low temperatures (log\textit{T} [K] $<$ 4.0), corresponding to the scaled-solar composition given either by \citet{grev98} or by \citet{palm14} (hereafter GS98 and PLJ14, respectively). Different choices of $(Z/X)_\sun$ correspond to different metal distributions in the solar structure, which, in their turn, change the calculated depth of the convective zone. Indeed, it was pointed out that SSMs with low metal abundances (i.e. with low $(Z/X)_\sun$ values) disagree with the helioseismologically measured sound speed, the depth of the convective zone, and the surface helium abundance \citep[see e.g.][]{bahc04}. Solving this disagreement, known as the ``solar abundance problem'', is an issue not related to $^7$Be decay and is therefore beyond the scope of this work. Here we show that the effects of using the new rate are independent from the solar mixture assumed and can be stated in a quite general way.

Finally, we have to mention that all the analyses presented in the various cases of this work have been performed by keeping all the physical parameters fixed, except for the $^7$Be electron-capture rate, to evaluate the specific role of this rate and to minimize the effects related to other inputs. The results obtained with the updated estimate of the $^7$Be electron-capture rate given by STPB13 have been compared with those obtained with the evaluation given by ADE11 for the two mentioned stellar choices of the chemical composition. 
In principle, different assumptions for the composition, i.e. for the metal abundances, may lead to differences in the solar core temperature, hence also  in the solar structure and in neutrino fluxes: see Section \ref{sec:fluxes} for a quantitative discussion. 
\begin{table} [t!]
\begin{center}
\caption{Major reaction rates included in the Standard Solar Models presented in this paper.} 
\label{tab:rates}
\begin{tabular}{c c}
\hline
\hline
 Reaction & Reference \\
\hline
 $^{1}$H(p, $\beta^+\nu_{\rm e}$)$^2$H  & 1 \\
 $^{1}$H(e$^-$p, $\nu_{\rm e}$)$^2$H & 2 \\
 $^{2}$H(p, $\gamma$)$^3$He & 2 \\
 $^{3}$He(p, $\beta^+\nu_{\rm e}$)$^4$He & 2 \\
 $^{3}$He($^{3}$He, $\alpha$)2H & 2 \\
 $^{3}$He($\alpha$, $\gamma$)$^7$Be & 2 \\
 $^{7}$Li(p, $\alpha$)$^4$He & 3 \\
 $^{7}$Be(p, $\gamma$)$^8$B & 4 \\
 $^{7}$Be(e$^-$, $\nu_{\rm e}$)$^7$Be & 2, 5 \\
 $^{12}$C(p, $\gamma$)$^{13}$N & 2 \\
 $^{13}$C(p, $\gamma$)$^{14}$N & 2 \\
 $^{14}$N(p, $\gamma$)$^{15}$O & 6 \\
 $^{15}$N(p, $\gamma$)$^{16}$O & 2 \\
 $^{15}$N(p, $\alpha$)$^{12}$C & 2 \\
 $^{16}$O(p, $\gamma$)$^{17}$F & 2 \\
 $^{17}$O(p, $\gamma$)$^{18}$F & 7 \\ 
 $^{17}$O(p, $\alpha$)$^{14}$N & 8 \\
 $^{14}$C(p, $\gamma$)$^{15}$N & 9 \\
 $^{18}$O(p, $\gamma$)$^{19}$F & 10 \\
 $^{18}$O(p, $\alpha$)$^{15}$N & 11 \\
 $^{19}$F(p, $\gamma$)$^{20}$Ne & 12 \\
 $^{19}$F(p, $\alpha$)$^{16}$O & 13 \\
 $^{6}$Li(p, $\gamma$)$^{7}$Be & 12 \\
 $^{6}$Li(p, $^3$He)$^{4}$He & 12 \\
 $^{9}$Be(p, $\gamma$)$^{10}$B & 12 \\
 $^{9}$B(p, $\alpha$)$^{6}$Li & 14 \\
 $^{10}$B(p, $\gamma$)$^{11}$C & 12 \\
 $^{10}$B(p, $\alpha$)$^{7}$Be & 14 \\
 $^{11}$B(p, $\gamma$)$^{12}$C & 12 \\
 $^{11}$B(p, $\alpha \alpha$)$^{4}$He & 12 \\
 $^{14}$C($\beta^-$, $\bar{\nu}_{\rm e}$)$^{14}$N & 15 \\ 
 $^{18}$F($\beta^+$, $\nu_{\rm e}$)$^{18}$O & 16 \\ 
 $^{18}$O($\beta^-$, $\bar{\nu}_{\rm e}$)$^{18}$F & 16 \\ 
\hline
\end{tabular}
\tablebib{
(1) \citet{marc13}; (2) \citet{adel11}; (3) \citet{lami12}; (4) \citet{zhan15}; (5) \citet{simo13}; (6) \citet{mart11}; (7) \citet{dile14}; (8) \citet{brun16}; (9) \citet{ilia10}; (10) \citet{buck12}; (11) \citet{laco10}; (12) \citet{angu99}; (13) \citet{inde17}; (14) \citet{lami15}; (15) \citet{raus00}; (16) \citet{oda94}.}

\end{center}
\end{table}
%
%
%
%
%
%
%
%
%
\section{Electron-capture on $^7$Be}
\label{sec:capture}
The deep stellar interiors are characterized by high densities and high temperatures. This implies that atoms are almost completely ionized; therefore, when describing the stellar core matter, it is necessary to apply the methods of plasma physics. The radioactive decay of a particular radioisotope (and its mean lifetime $\tau$) is strongly dependent, in such plasma conditions, on the density $\rho$ and the temperature $T$ of the plasma itself. In short, in order to provide an estimate of decay rates in stellar conditions one has to rely on accurate models for the plasma.

Many contributions, developed between the 60's and the 80's, considered a ionized plasma, whose degree of ionization is described through the Saha equation. Free electrons, acting as a screen inside the Debye radius, are treated as a Maxwellian gas \citep{taka87}. Concerning the specific case of $^7$Be electron-captures, the first detailed calculation from continuum states was done by \citet{bahc62}. Subsequently, estimates of the bound-electron contributions were also made \citep{iben67,bahc69,bahc94}. 
A recommended resulting rate, based on all these calculations, was proposed by \citet{adel98} and \citet{adel11}. More general treatments have also been developed over the years \citep{gruz97,brow97,sawy11}, but always referring to solar core conditions and maintaining an approach resembling the Born-Oppenheimer (BO) one. In addition to this, it was recognized that the major uncertainty affecting the decay rate arises from possible deviations from a pure Debye screening. Indeed, \citet{john92} estimated these possible corrections to the Debye-H\"uckel (DB) approximation by means of self-consistent thermal Hartree calculations, concluding that the  proposed rate was correct within an accuracy of about 2$\%$. 
In this regard, it has to be remarked that temperature at the 
centre of the Sun ($T\simeq 15.5 \, {\rm MK}$) is too high for electron degeneracy to set in. Hence, the classical approximation used e.g. by Bahcall to derive his rate is well founded for the solar conditions.

Quite recently \citet{simo13} developed a first-principles approach to derive the ${^{7}}$Be electron-capture rate, by modeling the electron-capture as a two-body scattering process ${^{7}}$Be-e${^{-}}$. To this aim, the e${^{-}}$-capture process is assumed to be proportional to the electronic density at the nucleus $\rho_e(0)$, which is screened and modified by the presence of the surrounding particles. 
We notice in passing that the DB approximation used by Bahcall represents the high-temperature classical limit of the approach developed by \citet{simo13}, which provides the e${^{-}}$-capture rate on ${^{7}}$Be over a range of plasma densities and temperatures definitively larger than that in the solar core conditions.

In this approach, the plasma is assumed hot and is modeled as a homogeneous Fermi gas made by ${^{7}}$Be atoms, surrounded by $N_p$ protons (hydrogen nuclei) and $N_e$ electrons, at various temperatures $T$ and densities $\rho$. The motion of quantum Fermi gases is ruled by the Schr\"odinger equation and described in a reference frame fixed on the Be nucleus. Due to the adopted non-inertial frame, the Hamiltonian of the system contains non-inertial terms, coupling the motion of particles of the different species. As Be is definitively more massive, all these terms can be safely neglected, so that a factorization of the eigenfunctions can be performed and separable eigensolutions can be found. This procedure is reminiscent of the conditions for the adiabatic theorem, and thus it represents an ``adiabatic'' approximation. In this way the many-body scattering problem is reduced to a screened two-body problem, so that $\rho_e(0)$ is computed by solving a coupled Hartree-Fock (HF) self-consistent system of equations for both protons and electrons, in the electric field generated by the ${^{7}}$Be nucleus located at the origin of the reference frame. The HF treatment of the Coulomb repulsion is satisfactory and accurate enough to comply with the electron correlation in stellar conditions \citep[see ][]{simo13}.

\begin{figure}[t!]  
\begin{center}
\includegraphics[width=\columnwidth]{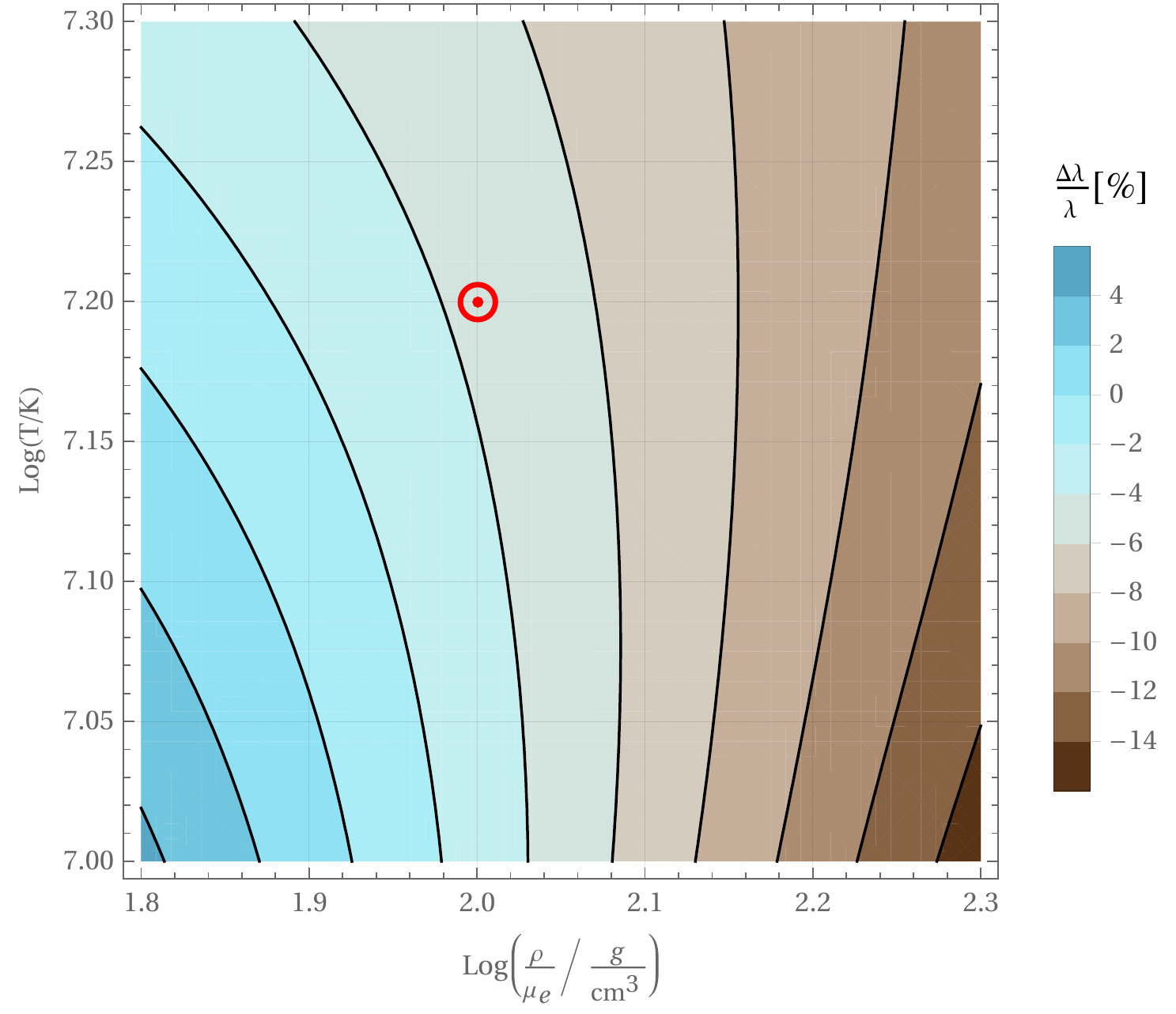}
\caption{The fractional variation of the $^7$Be electron-capture rate, $\Delta \lambda / \lambda \left[ \% \right]$ = $100\cdot(R_{\rm STPB13}-R_{\rm ADE11})/ R_{\rm ADE11}$, as a function of $\rho/{\mu_{\rm e}}$ and $T$, adopting the \citet{simo13} rate, as compared to the \citet{adel11} one, for the PLJ14 solar composition (see Section \ref{sec:ssm}). The solar core conditions are highlighted with the common solar symbol. 
A color version of this figure is available in the online journal.}
\label{fig:rates1}
\end{center}
\end{figure}

The mean lifetime, resulting from this method, is in general compatible with estimates by \citet{bahc62,bahc69,bahc94,adel98,adel11}; however, it has values that, in solar conditions, are smaller by $\sim 3-4\%$ with respect to those estimated in the mentioned works. Far from these conditions, the differences can be much more pronounced (see Figure \ref{fig:rates1}).
We refer the reader to \citet{simo13} for the details of the calculations.
The total reaction rate $\lambda$ for $^7$Be(e$^-$,$\nu_{\rm e}$)$^7$Li by STPB13 can also be expressed analytically in an approximate formula, as a function of temperature, density, and composition. 

An expression that agrees with an accuracy of 2\% to the tabulated results for the rate $\lambda$ $[{\rm s}^{-1}]$, in the region of relevance for solar physics, i.e. 
$35 \lesssim \rho/\mu_e$ ${\rm [g \, cm^{-3}]} \lesssim 105$ and $10\le T_6$ ${\rm [MK]} \le 16$, is:
\begin{equation}\label{eq:formula}
\lambda(\frac{\rho}{\mu_{\rm e}},T_6)= \frac{\rho}{\mu_{\rm e}} \frac{\kappa}{\sqrt{T_{6}}} \Big[ 1+ \alpha\, (T_6-16) + \beta\, \frac{\rho}{\mu_{\rm e}} \, \Big(1 + \gamma\, (T_6-16)\Big)  \Big] \, .
\end{equation}
Here $\mu_e$ is the mean molecular weight per electron, $T_6$ is the temperature in units of $10^6$ K, and $\rho$ is the density in units of ${\rm [g \, cm^{-3}]}$.
Thus, the electron density is $n_e$ = $\rho /(m_p \mu_e)$, where $m_p$ is the proton mass.
The values of the four coefficients $\kappa, \alpha, \beta,\gamma$, whose units ensure the correct dimension of Eq. (\ref{eq:formula}), are reported in Table \ref{tab:coef}. We notice that a non-linear term in the density is present, while it was absent in Bahcall's calculations. In fact, this term is due to the Coulomb repulsion (electron screening) acted upon the electrons, which modifies the density close to the nucleus. Taking into account such a non-linearity requires the introduction of a higher number of polynomial terms.
We recall, however, that in this work we make use of a tabulated version of the decay rate by STPB13: in fact, the adopted fine resolution allows us to compute highly accurate solar models without adding further uncertainties deriving from the use of an analytical formula. 
Notice that in our discussion, none of the nuclear reaction rates relevant for the standard solar model has been modified, so that expected variations are entirely due to the new approach adopted in computing $^7$Be electron-capture rate. Nevertheless, the change in the electron density, due to the formalism introduced by \citet{simo13} to describe e$^-$-capture on $^7$Be might be relevant also for other charged-particle interactions, leading to a correction in the screening factor. An investigation of this possibility and the quantitative estimation of this effect deserves dedicated analyses and future work.
%
\begin{table*}\label{tab:coef}
\begin{center}
\caption{Coefficients for the analytical approximation to the STPB13 and ADE11 electron-capture rates.}
\begin{tabular}{c c c c c}
& $\kappa$ & $\alpha$ & $\beta$ & $\gamma$ \\ \hline \hline
this paper & 
$ 5.9065\times 10^{-9}$  & 
$ -1.3614\times 10^{-2}$ & 
$ -9.2042\times 10^{-4}$ & 
$ -1.5334\times 10^{-1}$ \\ \hline
ADE11 & $5.6 \times10^{-9}$  & $+4\times 10^{-3}$ &$  0$ & $ 0$  \\ \hline
\end{tabular}
\end{center}
\end{table*}
\section{Solar Neutrino Fluxes}
\label{sec:fluxes}
Stars with initial mass $M \lesssim$ 1.2 $M_{\sun}$ primarily burn hydrogen through the pp-chain. The latter has three main branches, namely the ppI-, ppII-, and ppIII-cycles. 
The pp, $^8$B $\beta$-decay and hep reactions produce neutrino spectra with characteristic shapes and with energies from zero up to a maximum energy $q$. In particular, the neutrinos coming from the weak hep branch are the most energetic ones produced by the Sun ($q$ $\leq$ 18.773 MeV) and, thus, are observed in the SNO and Super-Kamiokande event distributions because they populate energy bins above the $^8$B neutrino endpoint. The electron-capture reactions p + e$^-$ + p and $^7{\rm Be}+{\rm e}^{-}$ produce, on the contrary, emission lines, possibly broadened by thermal effects. Concerning the $^7$Be neutrinos, they form two distinct lines, corresponding to population of both the ground state (89.5$\%$) and the first excited state (10.5$\%$) in $^7$Li \citep{viss18}. 

The ppI, ppII, and ppIII contributions to solar energy generation can be determined from measurements of the pp/pep, $^7$Be, and $^8$B neutrino fluxes. Being the relative rates very sensitive to the solar core temperature $T_c$, one can infer from neutrino fluxes important information about the physics of the solar interior.  Nowadays the pp, $^7$Be and $^8$B fluxes are quite well known, while the measured pep neutrino flux is strongly model-dependent. In particular, it depends on the metallicity assumed for estimating the competing CNO neutrinos \citep{agos18}. The solar core physics is sensitive to metallicity effects because of the free-bound/bound-free transitions in metals, which are important contributors to the opacity. This means that metallicity variations alter the solar core temperature and, in turn, the fluxes of temperature-sensitive neutrinos, such as those from $^8$B $\beta$-decay. 
Heavier metals (Mg, Si, and Fe) also affect the predicted neutrino fluxes \citep[see][]{bahc82}. Even if not very abundant, they are important opacity sources at the Sun center, as they are highly ionized. Instead, in the region just below the convective zone, at temperatures of a few millions kelvins, they are small contributors to the opacity. On the contrary, abundant, lighter, volatile heavy elemements (C, N, O, Ne, and Ar) are partially ionized there and significantly affect the radiative opacities. This is the origin of discrepancies between helioseismological measurements and the predictions made using solar compositions with low (Z/X), as discussed in \citet{bahc05b,bahc05c}. As a matter of fact, abundance variations of different metals influence different regions in the solar interior. Moreover, different CNO abundances imply an effect also on CNO burning efficiency (and corresponding neutrino fluxes) and a minor effect on the mean molecular weight and, in turn, on the thermodynamical quantities.

The net effect is that models using the GS98 compilation of abundances exhibit higher temperatures and higher densities with respect to those using the PLJ14 one (see Table \ref{tab:models}).
On the other hand, while pp and pep fluxes are only slightly modified, $^7$Be, $^8$B, $^{13}$N, $^{15}$O, and $^{17}$F neutrino fluxes are rather enhanced. Their fluxes are indeed strongly dependent on the central temperature $T_{\rm c}$, with a power law of the form $\Phi \propto T_{\rm c}^m$, with $m$ = 10.0, 24.0, 24.4, 27.1 and 27.8, respectively \citep[see][]{bahc96}.
CNO neutrino fluxes are enhanced also due to the increased burning efficiency caused by the higher CNO abundances in the GS98 compilation.
As was already mentioned, using modern solar compositions like the PLJ14 one, with low surface metal abundances, one gets solar models in disagreement with helioseismological measurements \citep[see][]{bahc04,basu04,bahc05a,sere11,haxt13,viny17}. 
We have checked that the predicted sound speed profiles of our computed SSMs are in agreement with others in the literature. We found that for the PLJ14 abundance choice the prediction disagrees with the measured one \citep{scho98}. Instead, the choice of the older GS98 composition gives a better match.

We recall however that this work is not aimed at giving the best prediction for the total neutrino fluxes nor at
probing the solar metallicity problem; rather, we want to probe the effects induced on solar neutrino fluxes by varying the $^7$Be electron-capture rate only, in the light of the mentioned evaluation by STPB13.
\begin{table}[t!]
\begin{center}
\caption{The main relevant quantities for the solar models adopting the ADE11 rate, as defined in the text. The models using the STPB13 rate show negligible variations for the same quantities. Here $R_{\rm CE}$ is the radius at the base of the convective envelope, $T_{\rm c}$ and $\rho_{\rm c}$ are the central temperature and density, $ \alpha_{MLT}$ is the value of the mixing-length parameter. $X_{\rm ini}$, $Y_{\rm ini}$, $Z_{\rm ini}$ and $(Z/X)_{\rm ini}$ are the initial hydrogen, helium and metal abundances by mass and the initial metal-to-hydrogen ratio, while $X_{\sun}$, $Y_{\sun}$, $Z_{\sun}$ and $(Z/X)_{\sun}$ are the corresponding present-day photospheric values.}
\label{tab:models}
\begin{tabular}{c c c}
\hline
\hline
 & GS98 & PLJ14\\
\hline
$R_{\rm CE}/R_{\sun}$ & 0.71628 & 0.72294\\
$T_{\rm c}$ $[10^7 \rm{K}]$ & 1.55031 & 1.54286\\ 
$\rho_{\rm c}$ ${\rm [g \, cm^{-3}]}$ & 149.377 & 148.325\\
$\alpha_{\rm MLT}$ & 2.31832 & 2.30317\\
$X_{\rm ini}$ & 0.70428 & 0.71092\\
$Y_{\rm ini}$ & 0.27703 & 0.27256\\
$Z_{\rm ini}$ & 0.01868 & 0.01653\\
$(Z/X)_{\rm ini}$ & 0.02653 & 0.02325\\
$X_{\sun}$ & 0.73656 & 0.74412\\
$Y_{\sun}$ & 0.24656 & 0.24103\\
$Z_{\sun}$ & 0.01688 & 0.01485\\
$(Z/X)_{\sun}$ & 0.02292 & 0.01995\\
\hline
\end{tabular}
\end{center}
\end{table}
%
%
\section{Impact of a revised $^7$Be + e$^-$ on the $^8$B neutrino flux}
\label{sec:impact}
\begin{figure}[t!]  
\begin{center}
\includegraphics[width=\columnwidth]{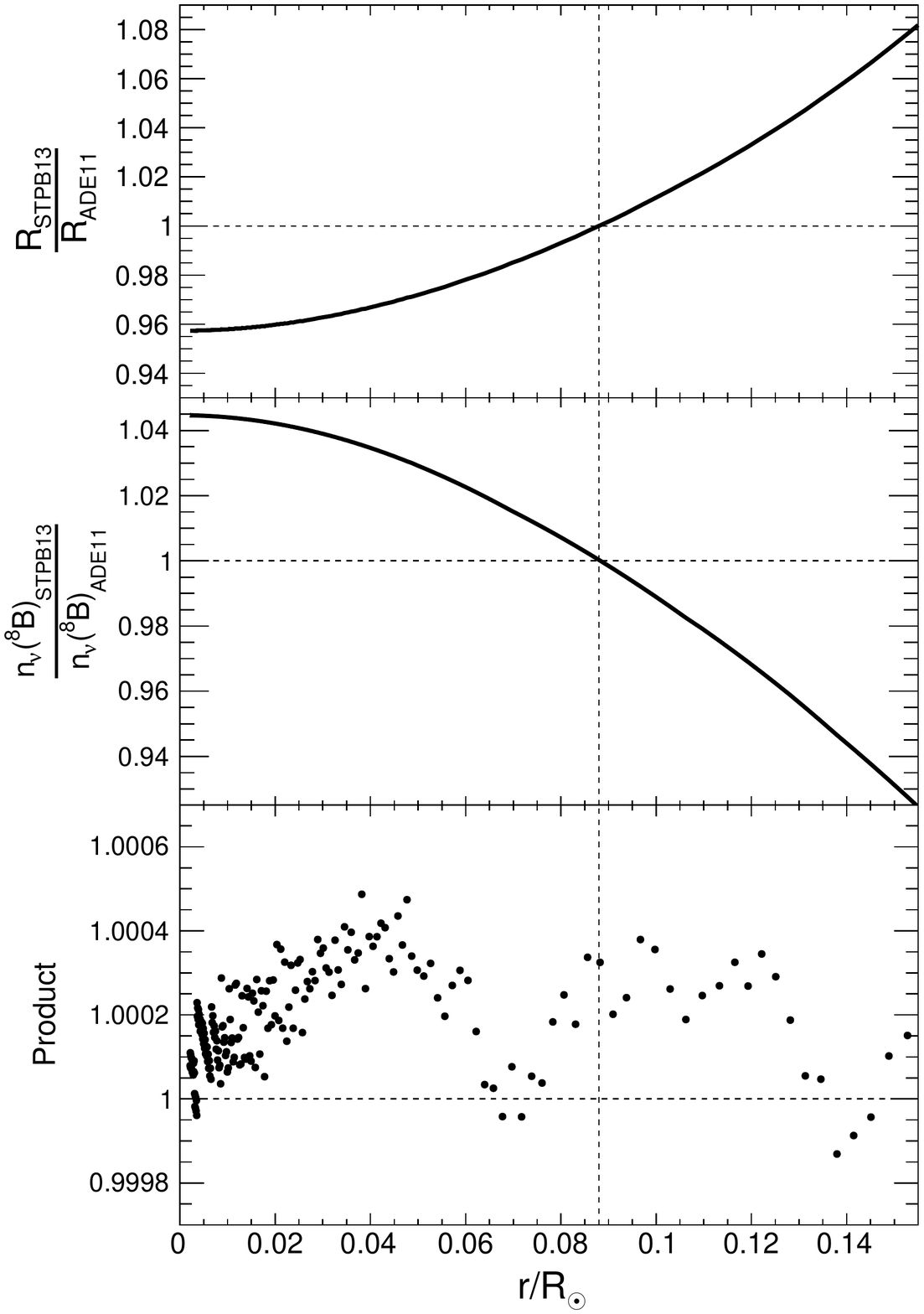}
\caption{Top panel shows the ratio between the STPB13 electron-capture rate and ADE11's one in the production region of $^8$B neutrinos, both computed on the solar structure resulting from the ADE11 SSM, with a PLJ14 composition. Middle panel shows the ratio between the neutrinos fraction produced in STPB13 SSM and a ADE11 one, both computed with a PLJ14 composition.
On the bottom panel the product $n_{\nu}(^8{\rm B})_{\rm STPB13} \cdot R_{\rm STPB13} / \left( n_{\nu}(^8{\rm B})_{\rm ADE11} \cdot R_{\rm STPB13} \right) $ is shown; note, in  comparison with the other two panels, the much finer vertical scale. The consistency of this value with the unity means that there is practically no difference in computing a SSM with the revised STPB13 rate or apply it directly on the solar structure of a ADE11 SSM.
}
\label{fig:rates2}
\end{center}
\end{figure}

In this section we want to evaluate the impact of using a revised rate for the $^7$Be electron-capture, computed following the approach suggested by \citet{simo13}, on the $^8$B neutrino flux. 
While pp neutrinos 
originate in a wide range of the Sun, corresponding to the main energy-producing region, $^7$Be and $^8$B neutrinos are produced in a hotter and narrower zone, ranging from the solar centre to about 0.15-0.2 R$_\sun$.
The quantities $R_{\rm STPB13}$ and $R_{\rm ADE11}$ represent the electron-capture rate given by STPB13 and by ADE11, respectively. 
The top panel of Fig. \ref{fig:rates2} shows the ratio between the STPB13 decay rate and ADE11's one in the production region of $^8$B neutrinos, both computed on the solar structure resulting from the ADE11 SSM, with a PLJ14 composition. As shown, there is an appreciable variation: the new rate is lower with respect to the ADE11 choice in solar core conditions, meaning that the $^7$Be neutrino production channel is slightly suppressed 
in favor of all other channels.
In particular, both the solar neutrino fluxes from $^7$Be and $^8$B,  $\Phi(^7{\rm Be})$ and $\Phi(^8{\rm B})$, are proportional to the local density of $^7$Be ions. The $\Phi(^7{\rm Be})$ flux depends on both the electron-capture ($R_{\rm ec}$) and the proton-capture rate ($R_{\rm pc}$) through:
\begin{equation}
\Phi(^7{\rm Be}) \propto \frac{R_{\rm ec}}{R_{\rm ec}+R_{\rm pc}} \; ,
\end{equation}
with $R_{\rm pc}$ $\approx$ 10$^{-3}$ $R_{\rm ec}$ \citep[see][]{adel98}. The flux  $\Phi(^7{\rm Be})$ is therefore basically independent from the rates and dependent only upon the branching ratio between the reactions $^{3}$He+$^{3}$He e $^{3}$He+$^{4}$He.
On the contrary, $\Phi(^8{\rm B})$ can be written as:
\begin{equation}\label{eq:ratio}
\Phi(^8{\rm B}) \propto \frac{R_{\rm pc}}{R_{\rm ec}+R_{\rm pc}} \simeq \frac{R_{\rm pc}}{R_{\rm ec}} \; ,
\end{equation}
meaning that it is inversely proportional to the electron-capture rate $R_{\rm ec}$. This means that a variation of the $R_{\rm ec}$ should have a linear effect on neutrino flux of $^8$B and negligible effects on other channels. Indeed, the STPB13 models present exactly the same physical and chemical features of the ADE11 models (see Table \ref{tab:models}).
If we take into account neutrinos that originate in each fraction of the solar radius (Figure \ref{fig:rates2}, middle panel), we thus deduce that, due to the less efficient electron-capture on $^7$Be rate, the $^8$B neutrino production channel becomes more efficient and so $\Phi(^8{\rm B})$ is increased.
It is also possible to see that, in correspondence of a change from negative to positive values of the variations in the electron-capture rate, the neutrino flux variation shifts from positive to negative values, thus corroborating the hypothesis of linearity between the electron-capture rate on $^7$Be and  the $^8$B neutrino flux. 
Furthermore, if relation (\ref{eq:ratio}) holds, then we see that:
\begin{equation}
\dfrac{n_{\nu}(^8{\rm B})_{\rm STPB13}}{n_{\nu}(^8{\rm B})_{\rm ADE11}} = \dfrac{\Phi(^8{\rm B})_{\rm STPB13}}{\Phi(^8{\rm B})_{\rm ADE11}}  \simeq \dfrac{R_{\rm ADE11}}{R_{\rm STPB13}} \; ,
\end{equation}
or, alternatively,
\begin{equation}\label{eq:one}
\dfrac{n_{\nu}(^8{\rm B})_{\rm STPB13}}{n_{\nu}(^8{\rm B})_{\rm ADE11}} \dfrac{R_{\rm STPB13}}{R_{\rm ADE11}} \simeq 1 \; ,
\end{equation}
where $n_{\nu}(^8{\rm B})$ is the number of neutrinos coming from the $^8{\rm B}$ decay. Bottom panel of Fig. \ref{fig:rates2} shows the product in the left-hand side of relation (\ref{eq:one}). Its value is consistent with unity at the sub-per mill level, meaning that relation (\ref{eq:ratio}) is indeed valid and that an increase of the $R_{\rm ec}$ has the effect of linearly decreasing the flux of $^8$B neutrinos. 
Finally, variations by +2.6$\%$ and +2.7$\%$ in $\Phi(^8{\rm B})$ are obtained for SSMs, using a PLJ14 or a GS98 compositions, respectively (see Table \ref{tab:fluxes}). The adoption of the STPB13 rate for electron-captures on $^7$Be has negligible effects on all other neutrino fluxes, because it induces no variation on the physics and the chemistry  of the SSM itself (see Table \ref{tab:models}).
\begin{table*}[t!!]
\begin{center}
\caption{This table presents the predicted fluxes, in units of 10$^{10}$ (pp), 10$^{9}$ ($^7$Be), 10$^{8}$ (pep, $^{13}$N, $^{15}$O), 10$^{6}$ ($^{8}$B, $^{17}$F), and 10$^{3}$ (hep) cm$^{-2}$s$^{-1}$ for the reference ADE11 models, presented in Table \ref{tab:models}, for the STPB13 models and relative differences.
.}
\begin{tabular}{c c c c c c c}
\hline
\hline
& \multicolumn{2}{c}{GS98} & & \multicolumn{2}{c}{PLJ14} & \\
\cline{2-3} \cline{5-6} 
& ADE11 & STPB13 & relative & ADE11 & STPB13 & relative\\
& & & differences & & & differences \\
\hline 
$\Phi$(pp)& 5.99 & 5.99 & +0.20\textperthousand & 6.01 & 6.01 & +0.01\textperthousand \\ 
$\Phi$(pep)& 1.42 & 1.42 & +0.25\textperthousand & 1.43  & 1.43 & +0.01\textperthousand \\
$\Phi$(hep)& 8.09 & 8.09 & +0.15\textperthousand & 8.22 & 8.22 &+0.01\textperthousand \\
$\Phi$($^7$Be)& 4.74 & 4.74 & +0.38\textperthousand & 4.54 & 4.54 & -0.01\textperthousand \\
$\Phi$($^{8}$B)& 5.28 & 5.42 & +2.70\% & 4.82 & 4.95 & +2.60\% \\
$\Phi$($^{13}$N)& 2.82 & 2.82 & +0.67\textperthousand & 2.55 & 2.55 & +0.06\textperthousand\\
$\Phi$($^{15}$O)& 2.07 & 2.07 & +0.71\textperthousand & 1.82 & 1.82 & +0.07\textperthousand\\
$\Phi$($^{17}$F)& 5.35 & 5.35 & +0.80\textperthousand & 3.95 & 3.95 & +0.07\textperthousand\\
\hline
\end{tabular}
\label{tab:fluxes}
\end{center}
\end{table*}
\subsection{Comparison with Solar neutrino fluxes}
At the present moment we cannot tag our predicted fluxes with well defined uncertainty estimates: we should construct Monte Carlo (MC) simulations of SSMs in order to provide statistical errors to our results \citep[see][]{bahc06,sere11,viny17}. Still we can estimate these uncertainties starting from known literature. Concerning the predicted $^8$B neutrino flux, \citet{bahc06} found that the 1$\sigma$ theoretical uncertainty varies from 17\% to 11\%, depending on the adopted composition (see their Table 15 and Figure 6). Similar but smaller values were also found by \citet{sere11} and \citet{viny17}. Then we can choose, in a conservative way, the larger value of 17\% as our uncertainty on the predicted $^8$B neutrino flux. 
Similarly we can adopt an error of 10\% 1$\sigma$ on the $^7$Be neutrino flux, as predicted by \citet{bahc06}, which is the highest found in the literature.
We also use, as correlation coefficient of the $^7$Be -$^8$B neutrino fluxes, the one given by \citet{bahc06} for the GS98 composition. 
In this way we only give a rough, but still reliable, estimate of the uncertainties affecting our neutrino flux predictions, to be compared with the measured values.

\begin{figure}[t!]  
\begin{center}
\includegraphics[width=\columnwidth]{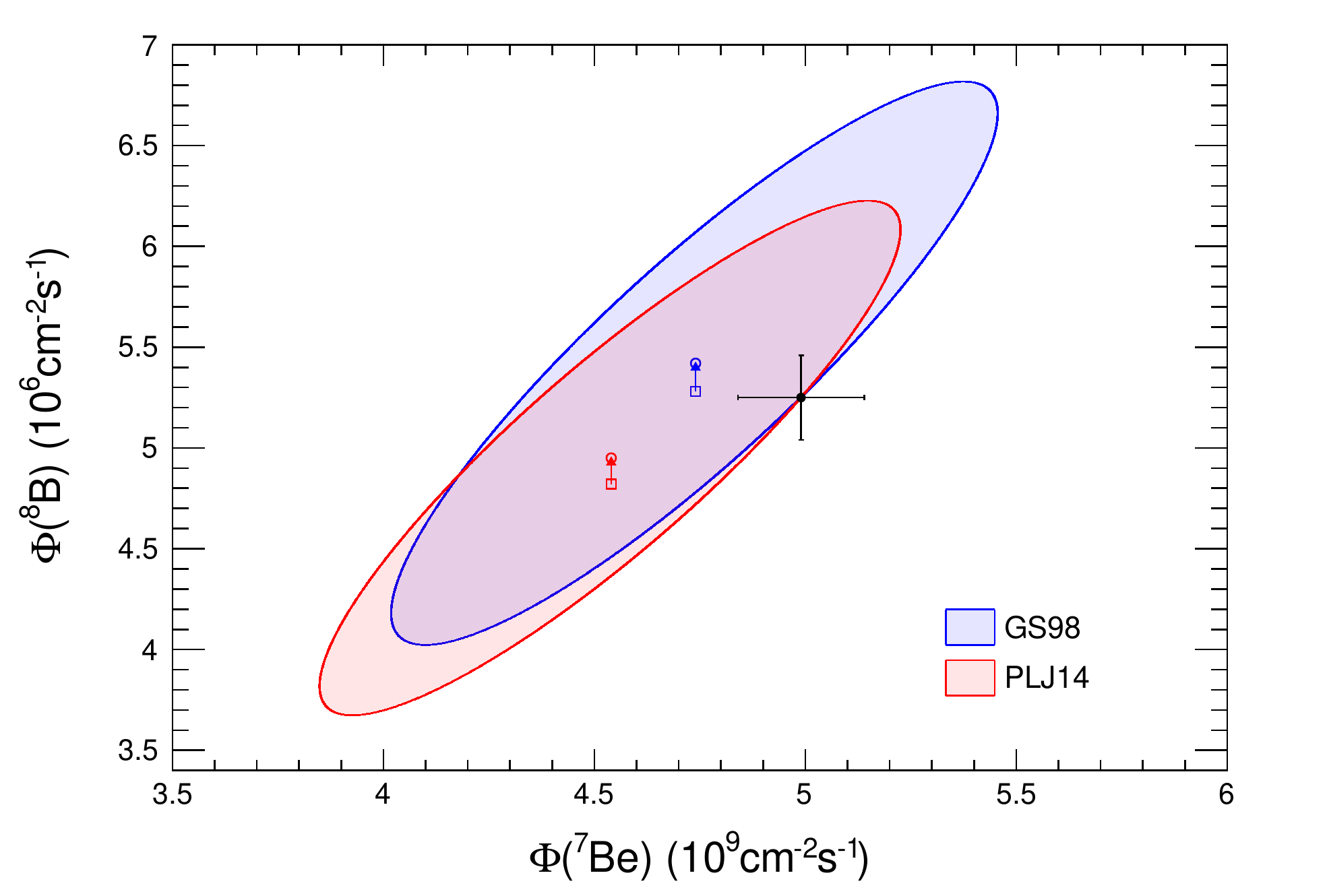}
\caption{$\Phi$($^{8}$B) and $\Phi$($^{7}$Be) fluxes compared to solar values \citep{ahar13,agos18}. Black dot and error bars indicate solar values, while squares and circles indicate the results obtained with the  ADE11 electron-capture rate (older) and the STPB13 (current) one, respectively. Ellipses denote theoretical 1$\sigma$ Confidence Level (C.L.) for 2 degrees of freedom.
A color version of this figure is available in the online journal.}
\label{fig:comparative}
\end{center}
\end{figure}

The final joint fit to all SNO data gave a total flux of  neutrino from $^8$B decays in the Sun of $\Phi(^8{\rm B})$ = 5.25(1 $\pm$ 0.04) $\times$ 10$^6$ cm$^{-2}$s$^{-1}$ \citep{ahar13}. The latest results of the Borexino collaboration \citep{agos18} provided a total flux of $^7$Be neutrino flux of $\Phi(^7{\rm Be})$ = 4.99(1 $\pm$ 0.03) $\times$ 10$^9$ cm$^{-2}$s$^{-1}$. Such a value is somehow model-dependent, being obtained, from the measured rates, assuming a specific mechanism of neutrino oscillations \citep[see][for details]{agos18}. In fact, elastic scattering measurements, like the ones performed by Borexino, are mainly sensitive to $\nu_{\rm e}$ Charged-Current interactions. On the contrary, the Neutral-Current detection channel in SNO is sensitive to all neutrino flavours and so it is a direct model-independent observation of the $^8$B solar neutrino flux.
Figure \ref{fig:comparative} shows that adopting either the GS98 or the PLJ14 compositions, leads to a fair agreement with the total $^8$B neutrino flux measured by the SNO neutral current experiments. The use of the revised electron-capture rate $R_{\rm STPB13}$ increases the old values of the predicted $^8$B neutrino fluxes with respect to the measured value. 
The measured value of the $^8$B neutrino flux is compatible with the solar model predictions for each of the two adopted solar compositions.
\section{Conclusions}\label{sec:concl}
We have presented new SSMs for two different mixtures of solar abundances, GS98 and PLJ14. Simulations have been performed with the FUNS code suite. We used recent values for the cross sections in our nuclear reaction network. 
In particular, we adopt the e$^-$-capture rate on $^7$Be provided by \citet{simo13} based on a description of the physical conditions in the solar interior more accurate than previous works (eg. ADE11) and applicable also for more general stellar environments.
A tabulated version of this rate is available in the online material.
The comparison with models computed with the ADE11 widely adopted electron-capture rate shows maximum differences of about 3-4\% in solar conditions. The effects on the standard solar model calculations, along with the effects on neutrino fluxes, have been discussed. We found that variations in the Solar structure and in neutrino fluxes are negligible, except for the $^8$B neutrino flux. The estimated increase is 2.6-2.7\%, depending on the composition assumed. Finally, we have also shown that the solar $^8$B neutrino flux 
is reproduced rather well, both using the GS98 and the PLJ14 abundance sets.
\begin{acknowledgements}
We warmly thank the referee, S. Degl'Innocenti, for the insightful comments and suggestions that helped us to improve the manuscript.
\end{acknowledgements}

%
%

\bibliographystyle{aa} 
\bibliography{biblio1}

\end{document}